\documentclass[12pt]{article}
\usepackage{mathrsfs}
\usepackage{amsmath,amssymb}

\usepackage{amsfonts}
\usepackage{latexsym}
\usepackage{amsthm}
\usepackage{pictex}

\usepackage[T2A]{fontenc}     % внутренняя {{T2A}} кодировка {{TeX}}
\usepackage[cp1251]{inputenc} % кодировка - можно использовать {{[cp866] [koi8-r]}}
\usepackage{graphicx}

\def\const{{\mathrm{const}}} % Konstante
\def\qed{\hbox {\hskip 1pt \vrule width 4pt height 6pt depth 1.5pt
        \hskip 1pt}}% Beweisende
\def\const{constant}

\def\Re{{\rm Re\,}}
\def\Im{{\rm Im\,}}
\def\lb{\label}
\newtheorem{theorem}{Theorem}

\def\l{\lambda}

\def\n{\nu}

\def\vp{\varphi}

\def\Z{{\Bbb Z}}
\def\R{{\Bbb R}}

\def\qq{\quad}
\newcommand{\ma}{\begin{pmatrix}}
\newcommand{\am}{\end{pmatrix}}
\newcommand{\ca}{\begin{cases}}
\newcommand{\ac}{\end{cases}}

\let\geq\geqslant
\let\leq\leqslant
\def\ma{\left(\begin{array}{cc}}
\def\am{\end{array}\right)}

\let\geq\geqslant
\let\leq\leqslant
\def\[{\begin{equation}}
\def\]{\end{equation}}

\def\/{\over}

\def\Re{\mathop{\rm Re}\nolimits}
\def\Im{\mathop{\rm Im}\nolimits}

\def\const{\mathop{\rm const}\nolimits}

\begin{document}

\date{\today}

\title{Asymptotics of the resonances for a continuously stratified layer}

\author{Ivan Argatov
\begin{footnote}
{Institute of Mathematics and Physics, Aberystwyth University,
Ceredigion SY23 3BZ, Wales, UK, email: ivan.argatov@gmail.com}
\end{footnote}
\and
Alexei Iantchenko
\begin{footnote}
{Malm{\"o} h{\"o}gskola, Teknik och samh{\"a}lle, 205 06
  Malm{\"o}, Sweden, email: ai@mah.se }
\end{footnote}
}

\maketitle

\begin{abstract}
Ultrasound wave propagation in a nonhomogeneous linearly elastic layer of constant thickness is considered.  The resonances for the corresponding acoustic propagator are studied. It is shown that the distribution of the resonances depends on the smoothness of the coefficients. Namely, if the coefficients have jump discontinuities at the boundaries, then the resonances are asymptotically distributed along a straight line parallel to the real axis on the unphysical sheet of the complex frequency plane. In the contrary, if the coefficients are continuous,  then it is shown that the resonances are asymptotically distributed along a logarithmic curve.
 The spacing between two successive resonances turns out to be sensitive to articular cartilage degeneration. The application of the obtained results to ultrasound testing of articular cartilage is discussed.
\end{abstract}

%%%%%%%%%%%%%%%%%%%%%%%%%%%%%%%%%%%%%%%%%%%%
%% MAINMATTER
%%%%%%%%%%%%%%%%%%%%%%%%%%%%%%%%%%%%%%%%%%%%

\section{Introduction}

Development of minimally invasive measurement techniques for assessing the viability of articular cartilage
and determining its physical and biomechanical properties is a problem of sound practical importance in surgery. Articular cartilage is a complex heterogeneous tissue with a sophisticated internal architecture exhibiting nonlinear and non-elastic behavior under mechanical loading and physical influence.
Motivated by the need for detecting the early-stage degeneration of articular cartilage, a number of studies in the past two decades have reported on the use of an ultrasound technique for the quantitative evaluation of structural and functional properties of articular cartilage \cite{Agemura_et_al1990,Chiang_et_al1994}.

While a constant ultrasound speed was assumed in the majority of theoretical studies \cite{Toyras_et_al1999,Nieminen_et_al1999}, it is well-known that the physical and biomechanical properties of articular cartilage  vary significantly across the thickness of the cartilage layer.
Heterogeneous structural and compositional properties along the thickness cause the variation of the ultrasound speed in articular cartilage. The depth dependence of the ultrasound speed in articular cartilage was experimentally demonstrated in the study  \cite{Agemura_et_al1990} on the propagation of ultrasound through various sections of articular cartilage at different depths. Moreover, it was suggested \cite{Patil_et_al2004} that the depth dependence and anisotropy of the ultrasound speed in articular cartilage should be taken into account for using ultrasound in articular cartilage measurement.

From a mathematical point of view, the problem of ultrasound wave propagation in a one-dimensional medium is related to the scattering problem for the acoustic propagator \cite{Wilcox}.
Reflection of acoustic plane waves from a continuously stratified layer was studied in a number of papers \cite{Robins1990,Wes1994}. In the present study, following \cite{Iantchenko2006}, we consider the resonance spectrum. The resonances (scattering poles) were studied in
\cite{Regge1958} from a physicists point of view. For an introduction to the mathematical theory of resonances see the review paper
\cite{Zworski1999}.

It is well known \cite{Regge1958} that the resonances are associated to the peaks of the reflection coefficient. That is why, it is important to have a clear understanding of the resonance distribution at high frequencies. Based on the asymptotic analysis, we show that the location of the resonances can be useful in developing ultrasound technique for assessment of articular cartilage degeneration.

\section{Physical problem formulation}

Consider an incident plane acoustic pressure wave of unit amplitude, propagating in the positive $z$ direction in a uniform fluid medium with density $\rho_0$ and sound speed $c_0$,
\begin{equation}
p_{\rm inc}^{(0)}(z,t)=e^{i k_0 z}e^{-i\omega t},
\label{0w(1.1)}
\end{equation}
where $k_0=\omega/c_0$ is the acoustic wavenumber.

The acoustic pressure in the upper medium satisfies the wave equation
\begin{equation}
\frac{\partial^2 p^{(0)}}{\partial z^2}(z,t)-\frac{1}{c_0^2}
\frac{\partial^2 p^{(0)}}{\partial t^2}(z,t)=0,\quad z\in(-\infty,0).
\label{0w(1.3)}
\end{equation}

We assume that the incident acoustic wave coming from the half-space $z<0$ illuminates the surface $z=0$ of a continuously stratified elastic layer of thickness $h$, whose density $\rho(z)$ and Lam\'e's elastic constants $\lambda(z)$ and $\mu(z)$ vary continuously with depth, while the bottom surface of the layer $z=h$ is firmly attached to a homogeneous, isotropic and elastic half-space $z>h$.

The vertical displacement functions $u_3(z,t)$ and $u_3^{(1)}(z,t)$ of the stratified elastic layer ($0<z<h$) and the homogeneous elastic half-space ($z>h$), respectively, satisfy the following differential equations:
\begin{equation}
\frac{\partial }{\partial z}\Bigl(\chi(z)\frac{\partial u_3}{\partial z}(z,t)\Bigr)=
\rho(z)\frac{\partial^2 u_3}{\partial t^2}(z,t),\quad z\in(0,h),
\label{0w(1.4)}
\end{equation}
\begin{equation}
\frac{\partial^2 u_3^{(1)}}{\partial z^2}(z,t)-\frac{1}{c_1^2}
\frac{\partial^2 u_3^{(1)}}{\partial t^2}(z,t)=0,\quad z\in(h,+\infty).
\label{0w(1.5)}
\end{equation}
Here, $\chi(z)=\lambda(z)+2\mu(z)$ is the so-called aggregate elastic modulus of the layer,
$c_1^2=(\lambda_1+2\mu_1)/\rho_1$ with $\rho_1$ and $\lambda_1$, $\mu_1$ being the density and Lam\'e's elastic constants of the half-space $z>h$.

At the solid-solid interface $z=h$, the following boundary conditions take place:
\begin{equation}
u_3(h,t)=u_3^{(1)}(h,t),
\label{0w(1.6)}
\end{equation}
\begin{equation}
\chi(h)\frac{\partial u_3}{\partial z}(h,t)=(\lambda_1+2\mu_1)
\frac{\partial u_3^{(1)}}{\partial z}(h,t).
\label{0w(1.7)}
\end{equation}

At the fluid-solid interface $z=0$, the following boundary conditions should be satisfied:
\begin{equation}
u_3^{(0)}(0,t)=u_3(0,t),
\label{0w(1.8)}
\end{equation}
\begin{equation}
-p^{(0)}(0,t)=\chi(0)\frac{\partial u_3}{\partial z}(0,t).
\label{0w(1.9)}
\end{equation}
Here, $u_3^{(0)}(z,t)$ is the vertical displacement function in the upper medium, which is related to the acoustic pressure through the equation
\begin{equation}
\frac{\partial^2 u_3^{(0)}}{\partial t^2}(z,t)+\frac{1}{\rho_0}
\frac{\partial p^{(0)}}{\partial z}(z,t)=0,\quad z\in(-\infty,0).
\label{0w(1.10)}
\end{equation}

In what follows we assume that the time dependance of the solution to the problem (\ref{0w(1.1)})\,--\,(\ref{0w(1.10)}) is assumed to be of the form $\exp(-i\omega t)$.

\section{Resonances for the acoustic propagator}

The physical problem (\ref{0w(1.1)})\,--\,(\ref{0w(1.10)}) formulated in the previous section can be reduced to the following spectral problem:
\begin{equation}
-\frac{1}{\rho(z)}\frac{d }{d z}\Bigl(\chi(z)\frac{d U}{d z}(z)\Bigr)=\omega^2 U(z).
\label{0w(1.4B)}
\end{equation}
Here, $\rho(z)$ and $\chi(z)$ are real positive piecewise smooth functions with discontinuities at the ends of the interval  $(0,h)$ such that
$$
\chi(z)=\left\{\begin{array}{cc}
                \chi_0,  & z<0 \\
                 \chi_1, & z>h
               \end{array}\right.,\quad
\rho(z)=\left\{\begin{array}{cc}
                \rho_0,  & z<0 \\
                 \rho_1, & z>h
               \end{array}\right.,
$$
where $\chi_0$, $\chi_1$ and $\rho_0$, $\rho_1$ are positive constants. Note that $\omega$ in equation~(\ref{0w(1.4B)}) is now allowed to be a complex number.

Let us also introduce the following notation:
$$
c(z)=\sqrt{\frac{\chi(z)}{\rho(z)}},\quad
c_0=\sqrt{\frac{\chi_0}{\rho_0}},\quad
c_1=\sqrt{\frac{\chi_1}{\rho_1}},\quad
m(z)=\sqrt{\frac{c(z)}{\chi(z)}},\quad
m_0=\sqrt{\frac{c_0}{\chi_0}},\quad
m_1=\sqrt{\frac{c_1}{\chi_1}}.
$$

We are looking for continuous solutions $U(z)$ to  equation~(\ref{0w(1.4B)}) satisfying the following conditions:
$$
\chi_0\frac{d U}{d z}(0-)=\chi_-\frac{d U}{d z}(0+),\qquad
\chi_+\frac{d U}{d z}(h-)=\chi_1\frac{d U}{d z}(h+).
$$
Here, $\chi_-$ and $\chi_+$ are the limit values of $\chi(z)$ as $z\rightarrow 0+$  and $z\rightarrow h-$, respectively.

Following the standard approach \cite{Marchenko1986}, we introduce the Jost solutions $f^\pm(\omega,z)$ to equation~(\ref{0w(1.4B)}) such that
$$
f^-(\omega,z)=\exp\Bigl(-\frac{i\omega z}{c_0}\Bigr),\quad z< 0,\qquad
f^+(\omega,z)=\exp\Bigl(\frac{i\omega z}{c_1}\Bigr),\quad z>h.
$$

The resonances (scattering poles) are the complex roots of the generalized Wronskian $\{f^-,f^+\}$ considered on the unphysical sheet $\Im{\omega}< 0$ of the complex frequency plane (see \cite{Marchenko1986}, \cite{Iantchenko2012}). The Wronskian is defined as follows:
$$
\{f^-,f^+\}=\left\vert\begin{array}{cc}
                      f^-(\omega,z) & f^+(\omega,z) \\
             \displaystyle         \chi(z)\frac{d f^-}{d z}(\omega,z) & \displaystyle   \chi(z)\frac{d f^+}{d z}(\omega,z)
                    \end{array}
\right\vert.
$$

We apply an asymptotic method in the limit situation as $|\omega|\rightarrow\infty$. In order to construct the Jost solution $f^+(\omega,z)$, we make use of the Liouville transformation to reduce equation~(\ref{0w(1.4B)}) on the interval $(0,h)$ to the Schr{\"o}dinger form. After that we reformulate the differential equation in the form of Volterra integral equation taking into account the boundary conditions at the right end of the interval $(0,h)$.

An asymptotic representation for $f^+(\omega,z)$ is obtained by the method of successive iterations. Since the Wronskian $\{f^-,f^+\}$ is independent of $z$ and is a function of $\omega$ only, we can evaluate $\{f^-,f^+\}$ at the left end of the interval $(0,h)$, taking into account the corresponding boundary conditions. As a result, we obtain the asymptotic expansion of  the resonances in powers of $\omega.$
\section{Formulation of the main results}
Let us denote \begin{equation}
\tau=\int_0^h\frac{dz}{c(z)}.
\label{0w(R.2)}
\end{equation}
\begin{theorem}\lb{th-main}
The resonances for the acoustic propagator (\ref{0w(1.4B)})
are given by the following asymptotic formulas in the leading order, as $|\omega|\rightarrow\infty$:\\
i) If $m_+\neq m_1$ and $m_-\neq m_0$ (discontinuity  of the coefficients $\chi(z),$ $\rho(z)$), then
\begin{equation}
\omega_n \simeq -\frac{i}{2\tau}\ln\Xi+\frac{\pi n}{\tau},\quad n\in\Z,\quad \vert n\vert\to\infty,
\label{0w(R.1)}
\end{equation}
where
$$
\Xi=\frac{(m_0^2+m_-^2)(m_1^2+m_+^2)}{(m_0^2-m_-^2)(m_1^2-m_+^2)}.
$$

ii) If $m_+=m_1,$ $m_-=m_0,$ $m_+'\neq 0,$ $m_-'\neq 0$ (discontinuity of the derivative of the coefficients  $\chi(z),$ $\rho(z)$), then
\begin{equation}\label{discder}
\omega_n\simeq \frac{\pi n}{\tau}-i \left( \frac{\ln |2 \pi n|}{\tau}-\frac{\ln{\tau}\sqrt{\Theta}}{\tau}\right),\qq n\in\Z,\quad \vert n\vert\to\infty,
\end{equation}
where ${\displaystyle \Theta=m_0m_1\chi_0\chi_-m_-'m_+'   .}$

iii) If $m_+\equiv m(h-)=m_1,$ $ m_-\equiv m(0+)=m_0,$ $ m_+'=m_-'=0,$ $ m_+''\cdot m_-''> 0$
(discontinuity of the second derivative of the coefficients  $\chi(z),$ $\rho(z)$), then
\begin{equation}\label{discsecondder}
\omega_n\simeq\frac{\pi n}{\tau} -i\left( \frac{2}{\tau}\ln |2\pi n| -\frac{\ln\tau^2\sqrt{V(y_-)V(y_+)}}{\tau}   \right),\quad n\in\Z,\quad \vert n\vert\to\infty,
\end{equation}
where
$$ V(y+)=-\frac{c^2_1}{m_1}m_+'',\qq V(y-)=-\frac{c^2_0}{m_0}m_-''.$$
\end{theorem}
We prove the Theorem in Section \ref{s-proof}.

{\em Remarks.} 1) Note that in the case i) of discontinuous coefficients, the resonances $\omega_n$ are asymptotically distributed along the string parallel to the real axis on the  complex half-plane $\Im\omega<0$ (see formula (\ref{0w(R.1)})), while for the continuous coefficients in cases ii) and iii), the resonances $\omega_n$ are asymptotically distributed along the logarithmic curve with $\Im\omega_n\rightarrow-\infty.$ This behavior is typical for  the Schr{\"o}dinger operators with smooth
potentials on the line or even more generally for non-trapping $n-$dimensional scattering (see \cite{Zworski1999}).

2) It is to note that formula (\ref{0w(R.1)}) assumes the inequality $\Xi>0$. The analogous result holds true also in the case $\Xi<0$, see (\ref{sigma_neg}).

3) As it was proven by Grinberg \cite{Grinberg1991},
  the function $$\sqrt{\rho\chi}\equiv\frac{\chi}{c}=\frac{1}{m^2}$$ is uniquely defined by its reflection coefficient $R_-$ and
 the constant $$\sqrt{\rho_1\chi_1}\equiv\frac{\chi_1}{c_1}=\frac{1}{m_1^2}.$$
   In the same articles is shown how to extract some information on the ruptures of the functions $\rho,$ $\l,$ $\mu$ (the Lam{\'e} coefficients) from the asymptotics of the reflection coefficients only.
%Using the results of Grinberg, the first author showes in \cite{Iantchenko2012} how the function $m$ can be reconstructed from the set of all resonances and the %constant $m_1.$

\section{Calculation of resonances}\lb{s-proof}
In this section we prove Theorem \ref{th-main}. Instead of the interval $(0,h)$ it is more convenient for us to consider the symmetric  interval $(-h/2,h/2).$
The correspondence between two situations is straightforward.
\subsection{Discontinuous coefficients}
Here we prove formula (\ref{0w(R.1)}). We rewrite equation (\ref{0w(1.4B)}) using the functions $c$ and $\chi$ as follows:
 \begin{equation}
-\frac{c^2(z)}{\chi(z)}\frac{d }{d z}\Bigl(\chi(z)\frac{d U}{d z}(z)\Bigr)=\omega^2 U(z).
\label{0w(1.4B)bis}
\end{equation}
In the interval  $(-h/2,h/2),$ where the coefficients $c$ and $\chi$ are smooth,
we apply the Liouville transform (see \cite{Aktosun1993})
$$y=\int_0^z\frac{1}{c(\xi)}d\xi,\qq\vp(\omega,y)=\sqrt{\frac{\chi(z)}{c(z)}}U(\omega,z)=\frac{1}{m}U(\omega,z).$$
Then, equation (\ref{0w(1.4B)bis}) is transformed to the Schr{\"o}dinger equation
\begin{equation}\lb{Schr}
-\frac{d^2\vp}{dy^2} +V(y)\vp=\omega^2\vp
\end{equation}
with the potential %$$V(y)=-\sqrt[4]{\chi\rho}\left\{\frac{\chi^\prime}{\rho}\left(\frac{1}{\sqrt[4]\chi\rho}\right)^\prime+\frac{\chi}{\rho}\left(\frac{1}{\sqrt[4]{\chi\rho}}\right)''\right\}=
%-\sqrt{\frac{\chi}{c}}\left\{\frac{c^2\chi'}{\chi}\left(\sqrt{\frac{c}{\chi}}\right)'+c^2\left(\sqrt{\frac{c}{\chi}}\right)''\right\},
%$$
$$V(y)=-\frac{1}{m}\left\{\frac{c^2\chi'}{\chi}m'+c^2m''\right\}.$$
 Let us introduce the notation $$\displaystyle y_-=-\int_{-h/2}^0\frac{d\xi}{c(\xi)},\qq \displaystyle y_+=\int_0^{h/2}\frac{d\xi}{c(\xi)},\quad
 \tau:=y_+-y_-=\int_{-h/2}^{h/2}\frac{d\xi}{c(\xi)}. $$
 Note the difference of this definition of $\tau$ from that in (\ref{0w(R.2)}) which is due to our choice of the symmetrical interval $(-h/2,h/2).$

 We extend the function $y(z)$ from $(-h/2,h/2)$ into $\R$ by continuity as follows:
$$y(z)=\left\{\begin{array}{lcl}
             y_-+\frac{1}{c_0}\left(z+\frac{h}{2}\right) & \mbox{for} & z\leq -h/2, \\ \\
             \int_0^z\frac{1}{c(\xi)}d\xi &  \mbox{for} & z\in (-h/2,h/2), \\ \\
             y_++\frac{1}{c_1}\left(z-\frac{h}{2}\right) & \mbox{for} & z\geq h/2.
           \end{array}\right. $$
The potential $V(y)$ vanishes outside  the interval $[y_-,y_+].$ Using the relation $y'(z)=c^{-1}(z),$ we get
$$\frac{dU(z)}{dz}=m'(z)\vp(y)+\frac{m(z)}{c(z)}\frac{d\vp}{dy}.$$

We consider the boundary conditions at $y_-$
\begin{align*} &U(-\frac{h}{2}+)=U(-\frac{h}{2}-)\qq\Leftrightarrow \qq m_0\vp(y_--)=m_-\vp(y_-+),\\
&\chi_0\frac{d U}{d z}(-\frac{h}{2}-)=\chi_-\frac{d U}{d z}(-\frac{h}{2}+)\,\,\Leftrightarrow\,\,
\chi_0\frac{m_0}{c_0}\frac{d\vp}{dy}(y_--) -\chi_-m_-^\prime\frac{m_0}{m_-}\vp(y_--)=\chi_-\frac{m_-}{c_-}\frac{d\vp}{dy}(y_-+).
\end{align*}
At $y_+,$ we get
\begin{align*}&U(\frac{h}{2}-)=U(\frac{h}{2}+)\qq\Leftrightarrow \qq m_+\vp(y_+-)=m_1\vp(y_++),\\
 &\chi_+\frac{d U}{d z}(\frac{h}{2}-)=\chi_1\frac{d U}{d z}(\frac{h}{2}+)\,\,\Leftrightarrow\,\,
\chi_+\frac{m_+}{c_+}\frac{d\vp}{dy}(y_+-)=\chi_1\frac{m_1}{c_1}\frac{d\vp}{dy}(y_++) -\chi_+m_+^\prime\frac{m_1}{m_+}\vp(y_++).
\end{align*}

For $z<-h/2,$ we use $\displaystyle f^-(\omega,z)=\exp\Bigl(-\frac{i\omega z}{c_0}\Bigr),$  $z=c_0(y-y_-)-h/2,$ and get
\begin{align*}
&\vp^-(y)=\frac{1}{m_0}f^-\Bigl(c_0(y-y_-)-\frac{h}{2}\Bigl)= \frac{1}{m_0}\exp{\left(-i\omega\left(y-y_--\frac{h}{2c_0}\right)\right)},\\
&\frac{d\vp^-(y)}{dy}=\frac{-i\omega}{m_0}\exp{\left(-i\omega\left(y-y_--\frac{h}{2c_0}\right)\right)}.
\end{align*}

For $z>h/2,$ we use $\displaystyle f^+(\omega,z)=\exp\Bigl(\frac{i\omega z}{c_1}\Bigr),$  $z=c_1(y-y_+)+h/2,$ and obtain
\begin{align*}
&\vp^+(y)=\frac{1}{m_1}f^+\Bigl(c_1(y-y_+)+\frac{h}{2}\Bigl)= \frac{1}{m_1}\exp{\left(i\omega\left(\frac{h}{2c_1}-y_+\right)\right)}\exp{(i\omega y)},\\
&\frac{d\vp^+(y)}{dy}=\frac{i\omega}{m_1}f^+\Bigl(c_1(y-y_+)+\frac{h}{2}\Bigl)= \frac{i\omega}{m_1}\exp{\left(i\omega\left(\frac{h}{2c_1}-y_+\right)\right)}\exp{(i\omega y)}.
\end{align*}
Now we construct the function $\displaystyle\frac{1}{m(z)}f^+(\omega,z)=\vp^+(\omega,y),$ $y\in (y_-,y_+),$ satisfying the following boundary conditions at $y_+:$
\begin{align*}
& m_+\vp^+(y_+-)=\exp{\Bigl(i\omega\frac{h}{2c_1}\Bigl)},\\
& \chi_+\frac{m_+}{c_+}\frac{d\vp^+}{dy}(y_+-)=\left(\chi_1\frac{i\omega}{c_1} -\chi_+\frac{m_+^\prime}{m_+}\right)\exp{\Bigl(i\omega\frac{h}{2c_1}\Bigl)}.
\end{align*}
Note that $\chi_+\frac{m_+}{c_+}=\frac{1}{m_+},$ $\frac{\chi_1}{c_1}=\frac{1}{m_1^2}.$
Now we will have
\begin{align}
& \vp^+(y_+-)=\frac{1}{m_+}\exp{\Bigl(i\omega\frac{h}{2c_1}\Bigl)},\label{y+1}\\
& \frac{d\vp^+}{dy}(y_+-)=m_+\left(\frac{i\omega}{m_1^2} -\chi_+\frac{m_+^\prime}{m_+}\right)\exp{\Bigl(i\omega\frac{h}{2c_1}\Bigl)}\label{y+2}.
\end{align}
%{\em as on page 16 of Argatov's notes}.
Further, let us construct the general solution to equation (\ref{Schr}) for $y\in (y_-,y_+)$ considering the function  $\tilde{V}=V\vp$ to be a known function
$$\vp=\frac{1}{2i\omega}\int_{y_+}^y\left( e^{-i\omega (\eta-y)}-e^{i\omega (\eta-y)}\right)\tilde{V}(\eta)d\eta +C_1e^{-i\omega y}+C_2e^{i\omega y},$$ where $C_1,$ $C_2$ are obtained using the boundary conditions (\ref{y+1}), (\ref{y+2}). As a result, we get
$$C_1=N_1\exp{\left(i\omega\Bigl(y_++\frac{h}{2c_1}\Bigl)\right)},\qq C_2=N_2\exp{\left(-i\omega\Bigl(y_+-\frac{h}{2c_1}\Bigl)\right)},$$ where
\begin{equation}\lb{N1N2}
N_1=\frac12\left\{\frac{1}{m_+}-m_+\left(\frac{1}{m_1^2} -\frac{\chi_+m_+'}{i\omega m_+}\right)\right\},\qq N_2=\frac12\left\{\frac{1}{m_+}+m_+\left(\frac{1}{m_1^2} -\frac{\chi_+m_+'}{i\omega m_+}\right)\right\}.
\end{equation}

Now, we introduce a new unknown function
$X=e^{-i\omega \left(y-y_++\frac{h}{2c_1}\right)}\vp$ by solving the integral equation
$$X=N_1e^{i2\omega(y_+- y)} +N_2+\frac{1}{2i\omega}\int_{y_+}^y\left(1-e^{2i\omega(\eta-y)}\right)V(\eta)X(\eta)d\eta,\qq y\in (y_-,y_+), $$
where $N_1,$ $N_2$ are given in (\ref{N1N2}).

Then, at the left endpoint of the interval $[y_-,y_+],$ we obtain
\begin{equation}\lb{tobeiterated}
X(y_-)=N_1e^{i2\omega(y_+- y_-)} +N_2-\frac{1}{2i\omega}\int_{y_-}^{y_+}\left(1-e^{2i\omega(\eta-y_-)}\right)V(\eta)X(\eta)d\eta.
\end{equation}

For $\Im\omega\geq 0,$ it follows $|X|<\const$ (see \cite{Marchenko1986}).
Now let $\Im\omega <0.$ The first iteration of (\ref{tobeiterated}) gives
\begin{align*}
&X(y_-)=e^{-i\omega \left(y_--y_++\frac{h}{2c_1}\right)}\vp(y_-)\\
&= N_1e^{i2\omega\left(y_+-y_-\right)} +N_2-
-\frac{1}{2i\omega}\int_{y_-}^{y_+}\left(1-
e^{2i\omega(\eta -y_-)}\right)V(\eta)\left(N_1e^{i2\omega\left(y_+-\eta\right)}+N_2
\right)\\
&\phantom{2}\,\,+{\mathcal O}\left( |\omega|^{-3}e^{-2(\Im\omega)_-\tau}\right).
\end{align*}
By integrating by parts in the integral and putting $V_0=\int_{y_-}^{y_+}V(\eta)d\eta,$ we get
\begin{align*}& X(y_-)= e^{-i\omega \left(y_--y_++\frac{h}{2c_1}\right)}\vp(y_-)=\\
&N_1e^{i2\omega\left(y_+-y_-\right)} +N_2+\frac{1}{2i\omega}N_1e^{i2\omega(y_+-y_-)}V_0
-\frac{1}{(2i\omega)^2}e^{i2\omega(y_+-y_-)}(N_1V(y_-)-N_2V(y_+))\\
&-\frac{1}{2i\omega}N_2V_0+\frac{1}{(2i\omega)^2}(N_1V(y_+)-N_2V(y_-))+{\mathcal O}\left( |\omega|^{-3}e^{-2(\Im\omega)_-\tau}\right)
\end{align*}
and, consequently, \begin{align*}
\vp(y_-)=&\left(N_1+N_1\frac{V_0}{2i\omega}-\frac{1}{(2i\omega)^2}(N_1V(y_-)-N_2V(y_+))\right)e^{i\omega\left(y_+-y_-+\frac{h}{2c_1}\right)}\\
&+\left(N_2-\frac{1}{2i\omega}N_2V_0+\frac{1}{(2i\omega)^2}(N_1V(y_+)-N_2V(y_-))\right)e^{i\omega \left(y_--y_++\frac{h}{2c_1}\right)}\\
&+{\mathcal O}\left( |\omega|^{-3}e^{-(\Im\omega)_-(\tau+\frac12\tau_1)}\right).
\end{align*}

Now, we obtain in the leading order as $|\omega|\rightarrow\infty$
\begin{align*}
\frac{dX}{dy}=&N_1(-i2\omega)e^{i2\omega(y_+-y)}+\int_{y_+}^ye^{2i\omega (\eta -y)}V(\eta)X(\eta)d\eta \\
=&N_1(-i2\omega)e^{i2\omega(y_+-y)}-\int_y^{y+}e^{2i\omega (\eta -y)}V(\eta)\left( N_1e^{i2\omega(y_+-\eta)} +N_2\right)d\eta+\ldots
\end{align*} and
$$
\frac{dX}{dy}(y_-)=N_1(-i2\omega)e^{i2\omega(y_+-y_-)}-\int_{y_-}^{y+}e^{2i\omega (\eta -y_-)}V(\eta)\left( N_1e^{i2\omega(y_+-\eta)} +N_2\right)d\eta+\ldots .$$
Using the relations
$$\vp=e^{i\omega\left(y-y_++\frac{h}{2c_1}\right)}X,\qq\frac{d\vp}{dy}=e^{i\omega\left(y-y_++\frac{h}{2c_1}\right)}\left(i\omega X +\frac{dX}{dy}\right),$$
we get
\begin{align*}
\frac{d\vp}{dy}(y_-)=&e^{i\omega\left(y_--y_++\frac{h}{2c_1}\right)}\left(-i\omega N_1 e^{i2\omega(y_+-y_-)}+i\omega N_2 -\frac12 N_1V_0 e^{2i\omega(y_+-y_-)} -\frac12 N_2V_0\right.\\
&\left.-\frac12\int_{y_-}^{y_+}e^{2i\omega(\eta-y_-)}V(\eta)N_2d\eta -\frac12\int_{y_-}^{y_+}e^{2i\omega(y_+-\eta)}V(\eta)N_1d\eta\right)+\ldots.
\end{align*}
By integrating by parts we find
\begin{align*}
\frac{d\vp}{dy}(y_-)=&e^{i\omega\left(y_+-y_-+\frac{h}{2c_1}\right)}\left(-i\omega N_1 -\frac12 N_1 V_0 -\frac12\frac{V(y_+)N_2}{2i\omega}+ \frac12\frac{V(y_-)N_1}{2i\omega}\right)\\
+&e^{i\omega\left(y_--y_++\frac{h}{2c_1}\right)}\left(i\omega N_2 -\frac12 N_2 V_0 +\frac12\frac{V(y_-)N_2}{2i\omega}- \frac12\frac{V(y_+)N_1}{2i\omega}\right)+\ldots .
\end{align*}
Now we consider  the Wronskian
$$\{f^-,f^+\}(-h/2)=\left|\begin{array}{cc}
                            e^{i\omega \frac{h}{2c_0}} & m_-\vp^+(y_-) \\
                            -\chi_0\frac{i\omega}{c_0}e^{i\omega\frac{h}{2c_0}} & \chi_-m_-'(y_-)\vp^+(y_-) +\chi_-\frac{m_-}{c_-} \frac{d\vp^+}{dy}(y_-)
                          \end{array}
\right|.$$
Keeping only the coefficients not vanishing as $|\omega|\rightarrow \infty,$ we arrive at the asymptotic formula
\begin{align*}
&\chi_-m_-'(y_-)\vp^+(y_-) +\chi_-\frac{m_-}{c_-} \frac{d\vp^+}{dy}(y_-)\backsimeq\\
&e^{i\omega\left(y_+-y_-+\frac{h}{2c_1}\right)}\left\{\chi_-\frac{m_-}{c_-}(-i\omega)N_1 +\chi_-N_1\left(m_-'-\frac12\frac{m_-}{c_-}V_0\right)\right\}\\
&+e^{i\omega\left(y_--y_++\frac{h}{2c_1}\right)}\left\{\chi_-\frac{m_-}{c_-}(i\omega)N_2 +\chi_-N_2\left(m_-'-\frac12\frac{m_-}{c_-}V_0\right)\right\}.
%&+{\mathcal O}\left( |\omega|^{-1}e^{-2(\Im\omega)_-\tau}\right).
\end{align*}

Correspondingly, for the Wronskian we have
\begin{align*}
&\{f^-,f^+\}(-h/2)\backsimeq \\
&e^{i\omega\left(y_+-y_-+\frac{h}{2c_1}+\frac{h}{2c_0}\right)}\left\{
i\omega m_- N_1\left[-\frac{\chi_-}{c_-}+\frac{\chi_0}{c_0}\right]+\chi_-N_1\left(m_-'-\frac12\frac{m_-}{c_-}V_0\right)+\chi_0\frac{m_-N_1}{2c_0}V_0\right\}\\
+&e^{i\omega\left(y_--y_++\frac{h}{2c_1}+\frac{h}{2c_0}\right)}\left\{
i\omega m_- N_2\left[\frac{\chi_-}{c_-}+\frac{\chi_0}{c_0}\right]+\chi_-N_2\left(m_-'-\frac12\frac{m_-}{c_-}V_0\right)-\chi_0\frac{m_-N_2}{2c_0}V_0\right\}.
\end{align*}
In the leading order we obtain
\begin{align*}
&\{f^-,f^+\}\backsimeq \frac{i\omega}{2}m_-m_+ e^{i\omega\frac{h}{2}\left(\frac{1}{c_1}+\frac{1}{c_0}\right)}\left\{e^{i\omega\tau}
\left(\frac{1}{m_+^2}-\frac{1}{m_1^2}\right)\left(-\frac{1}{m_-^2}+\frac{1}{m_0^2}\right) \right.\\
&\left.+ e^{-i\omega\tau}
\left(\frac{1}{m_+^2}+\frac{1}{m_1^2}\right)\left(\frac{1}{m_-^2}+\frac{1}{m_0^2}\right)\right\}.
\end{align*}

Now we pass to  calculation of the resonances.
Setting the Wronskian equal to zero, we get in the leading order
\begin{align*}
e^{2i\omega(y_+-y_-)}=&\frac{N_2\left(\frac{\chi_-}{c_-}+\frac{\chi_0}{c_0}\right)}{N_1\left(\frac{\chi_-}{c_-}-\frac{\chi_0}{c_0}\right)}
=\frac{\left(1+\left(\frac{m_+}{m_1}\right)^2\right)\left(1+\left(\frac{m_-}{m_0}\right)^2\right)}{\left(1-\left(\frac{m_+}{m_1}\right)^2\right)\left(1-\left(\frac{m_-}{m_0}\right)^2\right)}\\
=& \frac{(m_0^2+m_-^2)(m_1^2+m_+^2)}{(m_0^2-m_-^2)(m_1^2-m_+^2)}=:\Xi.
\end{align*}
Recall the notation $$\tau:=y_+-y_-=\int_0^{h/2}\frac{d\xi}{c(\xi)}+\int_{-h/2}^0\frac{d\xi}{c(\xi)}=\int_{-h/2}^{h/2}\frac{d\xi}{c(\xi)}. $$
We have $|\Xi|>1.$ If $\Xi >1,$ then
$$\omega_n\backsimeq-i\frac{\ln\Xi}{2\tau} +\frac{\pi n}{\tau},\qq\n\in\Z,\qq |n|\rightarrow\infty.$$
If $\Xi <-1,$ then
\begin{equation}\lb{sigma_neg}\omega_n\backsimeq-i\frac{\ln(-\Xi)}{2\tau} +\frac{\pi}{2\tau} +\frac{\pi n}{\tau},\qq\n\in\Z,\qq |n|\rightarrow\infty.
\end{equation}
The proof of (\ref{0w(R.1)}) in Theorem \ref{th-main} is finished.\hfill\qed
\vspace{1cm}

{\em Remark.} Note that from  our calculations it follows the  asymptotic representation of the Jost function at $y=y_-$ in the leading order as $|\omega|\rightarrow\infty,$
\begin{align}
f^+(\omega,-h/2)=&m_-\vp^+(y_-)\backsimeq e^{i\omega\left(y_+-y_-+\frac{h}{2c_1}\right)}\left\{m_-N_1+m_-N_1\frac{V_0}{2i\omega}\right\}\nonumber\\
& +
e^{i\omega\left(y_--y_++\frac{h}{2c_1}\right)}\left\{m_-N_2-m_-N_2\frac{V_0}{2i\omega}\right\}.\lb{Jostas}\end{align}

If $\chi\equiv 1,$ then equation (\ref{0w(1.4B)bis}) is reduced to $-c^2U''=\omega^2U$ and,
as in (\ref{Jostas}), we get for $-\pi/2\leq\arg \omega\leq -\epsilon<0,$
\begin{align*}f^+(\omega,z)=&\frac{\sqrt{c(z)}}{2}e^{i\omega\frac{h}{2c_1}}\left\{e^{i\omega\int_z^{h/2}c^{-1}(s)ds}\left(\frac{1}{\sqrt{c_+}} -\sqrt{c_+}\right) \right.\\
&+\left. e^{-i\omega\int_z^{h/2}c^{-1}(s)ds}\left(\frac{1}{\sqrt{c_+}} +\sqrt{c_+}\right)\right\}(1+o(1)).
\end{align*}
This formula was earlier obtained by Pekker (Shubov) (in  \cite{Pekker1980}, page 159).

\subsection{Continuous coefficients}
{\em Proof of (\ref{discder}).} Suppose that $$m_+=m_1,\qq m_-=m_0\qq m_+'\neq 0,\qq m_-'\neq 0.$$ Then $$N_1=\frac12\frac{\chi_+m_+'}{i\omega},\qq N_2=\frac{1}{m_+}-\frac12\frac{\chi_+m_+'}{i\omega},$$ and as before we get in the leading order as $|\omega|\rightarrow\infty$

\begin{align*}
\{f^-,f^+\}&\simeq\frac{2i\omega}{m_0m_+}e^{i\omega \left(\frac12(\tau_0+\tau_1)\right)}\left(e^{i\omega \tau }\frac{1}{(2i\omega)^2}m_0m_1\chi_0\chi_1m_-'m_+' +e^{-i\omega \tau }\right)\\
&\simeq e^{i\omega \left(\tau +\frac12(\tau_0+\tau_1)\right)}\frac{1}{2i\omega}\chi_-\chi_+m_-'m_+' +e^{i\omega \left(-\tau +\frac12(\tau_0+\tau_1)\right)}
\left(\frac{2i\omega}{m_0m_+}\right).
\end{align*}
  Then we get that the  resonances are asymptotically close to the solutions of the equation
$$\frac{e^{2i\omega\tau}}{(2i\omega)^2}\Theta +1=0,\qq \Theta=m_0m_1\chi_0\chi_-m_-'m_+'   ,$$ and in the leading order as $|\omega|\rightarrow\infty$
 are given by
$$\omega_n\simeq \frac{\pi n}{\tau}-i \left( \frac{\ln |2 \pi n|}{\tau}-\frac{\ln{\tau\sqrt{\Theta}}}{\tau}\right),\qq n\in\Z,\qq |n|\rightarrow\infty.$$\hfill\qed

{\em Proof of (\ref{discsecondder}).}
 Suppose $$m_+=m_1,\qq m_-=m_0,\qq m_+'=m_-'=0,\qq m_+''\neq 0,\qq m_-''\neq 0.$$ Then
 $N_1=0,$ $N_2=\frac{1}{m_1}.$

  Note the following general relations
$$\{f^-,f^+\}=\frac{1}{m_0m_1}e^{i\omega\left(-\tau +\frac12(\tau_0+\tau_1)\right)}\left(2i\omega m_1X(y_-)+m_1\frac{dX}{dy}(y_-)\right)$$
and $$2i\omega m_1X(y_-)+m_1\frac{dX}{dy}(y_-)=\left(e^{2i\omega(y-y_-)}m_1X(y)\right)'(y_-).$$
By straightforward calculations we obtain $$ m_1 e^{2i\omega(y-y_-)}X(y)=e^{i2\omega(y-y_-)}+\frac{1}{2i\omega}\int_{y_+}^y\left(e^{i2\omega(y-y_-)}-e^{i2\omega(\eta-y_-)}\right)V(\eta)m_1X(\eta)d\eta
$$ and
$$\left(e^{2i\omega(y-y_-)}m_1X(y)\right)'(y_-)=i2\omega+\int_{y_+}^{y_-}V(\eta)m_1X(\eta)d\eta.$$
Thus, we get in the leading order
\begin{align*} &\left(e^{2i\omega(y-y_-)}m_1X(y)\right)'(y_-)\simeq i2\omega -V_0+\frac{1}{2i\omega}\int_{y_+}^{y_-}V(\eta)\int_{y_+}^\eta\left( 1-e^{2i\omega(\eta_2-\eta)}\right)V(\eta_2)d\eta_2d\eta\\
\simeq& i2\omega -V_0+\frac{1}{2i\omega}\int_{y_+}^{y_-}V(\eta)\int_{y_+}^\eta V(\eta_2)d\eta_2d\eta-\frac{1}{(2i\omega)^2}\int_{y_+}^{y_-}V^2(\eta)d\eta\\
&+
\frac{1}{(2i\omega)^2}\int_{y_+}^{y_-}V(\eta)e^{2i\omega(y_+-\eta)}V(y_+)d\eta\\
\simeq& i2\omega -V_0 -\frac{1}{(2i\omega)^2}\int_{y_-}^{y_+}V(\eta)e^{2i\omega(y_+-\eta)}V(y_+)d\eta
\simeq i2\omega -V_0 - \frac{V(y_-)V(y_+)}{(2i\omega)^3}e^{2i\omega\tau}.
\end{align*}
Therefore,  we get
$$\{f^-,f^+\}\simeq \frac{1}{m_0m_1}e^{i\omega\left(-\tau +\frac12(\tau_0+\tau_1)\right)}\left(i2\omega -V_0-\frac{V(y_-)V(y_+)}{(2i\omega)^3}e^{2i\omega\tau}\right).
$$
The resonances are asymptotically close to the solutions of the equation
$$e^{2i\omega\tau}\frac{V(y_-)V(y_+)}{(2i\omega)^4} -1=0 $$
and are given in the leading order by the asymptotic formula
$$\omega_n\simeq\frac{\pi n}{\tau} -i\left( \frac{2}{\tau}\ln |2\pi n| -\frac{\ln\tau^2\sqrt{V(y_-)V(y_+)}}{\tau}   \right),\quad n\in\Z,\quad \vert n\vert\to\infty,$$
where
$$ V(y+)=-\frac{c^2_1}{m_1}m''(h/2-),\qq V(y-)=-\frac{c^2_0}{m_0}m''(-h/2+)$$ and $V(y_-)V(y_+)>0.$\hfill\qed

\section{Discussion}

First of all, observe that the parameter $\tau$ introduced by equation~(\ref{0w(R.2)}) has the dimension of time and is interpreted as the time needed for the ultrasound wave to pass through the articular cartilage thickness.

In ultrasound experiments, one can observe the Breit--Wigner peaks on the profile of the scattering phase
$$\sigma(\omega)=\frac{1}{2\pi i}\log\det S(\omega),$$ where $S$ is the scattering matrix. These peaks are produced by the resonances  $\omega_n$ on the complex plane located in the neighborhood of the real axis.
 We have the Breit-Wigner formula for an isolated resonance $\omega_n:$
\begin{equation}\label{BreitWigner}
\sigma'(\omega)\simeq-\frac{1}{\pi}\frac{\Im\omega_n}{|\omega-\omega_n|^2},\qq \omega\simeq\Re\omega_n.
\end{equation}

  The asymptotic formula (\ref{0w(R.1)}) gives approximate locations for these resonances. According to equation~(\ref{0w(R.1)}), the distance between the resonances measured along the real axis is asymptotically close to $\pi/\tau$.

Thus, from the experimentally observable distance $\Delta\omega$ between two successive peaks for the reflection coefficient, we can estimate the characteristic time $\tau$ as follows:
$$
\tau\approx \frac{\pi}{\Delta\omega}.
$$
We note that the Breit-Wigner formula (\ref{BreitWigner})  is  not really rigorous without further parameters (such as the semiclassical parameter) since we do not know the size of the contribution of the other terms (see \cite{PetkovZworski1999}).

Finally, from the mean value theorem for integrals, it follows that
$$
\tau=\frac{h}{c(z_*)},
$$
where $z_*$ is some point in the interval $(0,h)$.

It is known \cite{Kiviranta2007} that for some type of degeneration of articular cartilage, the cartilage thickness increases, while its elastic modulus decreases. So, the characteristic parameter  $\tau$ will increase with the degeneration. In other words, the degeneration of articular cartilage will increase the spacing between the resonances. This fact can be useful in developing laboratory tests for assessing articular cartilage viability.

%\endinput


\begin{thebibliography}{9}

\bibitem{Agemura_et_al1990}
D.H.~Agemura, W.D.~Jr.~O'Brien, J.E.~Olerud, L.E.~Chun, D.E.~Eyre,
\emph{J. Acoust. Soc. Am.} \textbf{87}, 1786--1791 (1990).

\bibitem{Aktosun1993}
T.~Aktosun,
\emph{J. Math. Phys.} \textbf{34} (5), 1619--1634 (1993).


\bibitem{Chiang_et_al1994}
E.H.~Chiang, R.S.~Adler, Ch.R.~Meyer, J.M.~Rubin, D.K.~Dedrick, T.J.~Laing,
\emph{Ultrasound in Medicine \& Biology} \textbf{20}, 123--135 (1994).





\bibitem{Grinberg1991}
N.I.~Grinberg,
\emph{Inverse Problems} \textbf{7}, 567--576 (1991).





\bibitem{Iantchenko2006}
A.~Iantchenko,
\emph{Appl. Analysis} \textbf{85}, 1383--1410 (2006).

%\bibitem{Iantchenko2012}
%A.~Iantchenko,
%\emph{Inverse resonance problem for elastic layered medium}, Preprint









\bibitem{Kiviranta2007}
P.~Kiviranta, J.~T\"{ц}yr\"{a}s, M.T.~Nieminen, M.S.~Laasanen, S.~Saarakkala, H.J.~Nieminen, M.J. Nissi, J.S.~Jurvelin,
\emph{Europ. Cells and Mater.} \textbf{13}, 46--55 (2007).

\bibitem{Marchenko1986}
V.~Marchenko,
\emph{Sturm--Liouville Operator and Applications}, Birkh\"{a}user, Basel, 1986.

\bibitem{Nieminen_et_al1999}
H.J.~Nieminen, J.~T$\ddot{\rm o}$yr$\ddot{\rm a}$s, J.~Rieppo, M.T.~Nieminen, J.~Hirvonen, R.~Korhonen, J.S.~Jurvelin,
\emph{Ultrasound in Medicine \& Biology} \textbf{28}, 519--525 (2002).


\bibitem{Pekker1980}
M.~Pekker,
\emph{Amer. Math. Soc. Transl.}  \textbf{115} (2), 143--164 (1980).

\bibitem{PetkovZworski1999}
V.~Petkov, M.~Zworski,
\emph{Commun. Math. Phys.}  \textbf{204}, 329--351 (1999).

\bibitem{Patil_et_al2004}
S.G.~Patil, Y.P.~Zheng, J.Y.~Wu, J.~Shi,
\emph{Ultrasound in Medicine \& Biology} \textbf{30}, 953--963 (2004).

\bibitem{Regge1958}
T.~Regge,
\emph{Nuovo Cimento} \textbf{8}, 671--679 (1958).

\bibitem{Robins1990}
A.J.~Robins,
\emph{J. Acoust. Soc. Am.} \textbf{87}, 1546--1552 (1990).


\bibitem{Toyras_et_al1999}
J.~T$\ddot{\rm o}$yr$\ddot{\rm a}$s, J.~Rieppo, M.T.~Nieminen, H.J.~Helminen, J.S.~Jurvelin,
\emph{Phys. Med. Biol.} \textbf{44}, 2723--2733 (1999).

\bibitem{Wes1994}
Z.~Wesolowski,
\emph{Acta Mechanica} \textbf{105}, 119--131 (1994).

\bibitem{Wilcox}
C.H.~Wilcox,
\emph{Sound Propagation in Stratified Fluids}, Springer-Verlag, Berlin,  1984.

\bibitem{Zworski1999}
M.~Zworski,
\emph{Notes of the AMS} \textbf{46}, 319--328 (1999).


\end{thebibliography}
\end{document}